\documentclass[twocolumn]{aastex63}
\newcommand{\kms}{\,km\,s$^{-1}$} 
\newcommand{\mgbfe}{$\rm Mgb/\langle Fe\rangle$}
\submitjournal{APJ}

\shorttitle{SFHs of redspirals}
\shortauthors{Zhou et al.}
\graphicspath{{./}{figures_dr15/}}

\begin{document}

\title{Star formation histories of massive red spiral galaxies in the local universe}

\email{Contact e-mail: szhou@tsinghua.edu.cn (SZ); cli2015@tsinghua.edu.cn (CL)}

\author[0000-0002-8999-6814]{Shuang Zhou}
\affiliation{Department of Astronomy, Tsinghua University, Beijing 100084, China}

\author[0000-0002-8711-8970]{Cheng Li}
\affiliation{Department of Astronomy, Tsinghua University, Beijing 100084, China}

\author{Cai-Na Hao}
\affiliation{Tianjin Astrophysics Center, Tianjin Normal University, Tianjin 300387, China}

\author{Rui Guo}
\affiliation{Tianjin Astrophysics Center, Tianjin Normal University, Tianjin 300387, China}

\author{Houjun Mo}
\affiliation{Department of Astronomy, University of Massachusetts Amherst, MA 01003, USA}

\author{Xiaoyang Xia}
\affiliation{Tianjin Astrophysics Center, Tianjin Normal University, Tianjin 300387, China}

\begin{abstract}
We investigate the star formation histories (SFHs) of massive red spiral galaxies with stellar mass $M_\ast>10^{10.5}M_\odot$, and make comparisons with blue spirals and red ellipticals of similar masses. We make use of the integral field spectroscopy from the SDSS-IV/DR15 MaNGA sample, and estimate spatially resolved SFHs and stellar population properties of each galaxy by applying a Bayesian spectral fitting code to the MaNGA spectra. We find that both red spirals and red ellipticals have experienced only one major star formation episode at early times, and the result is independent of the adopted SFH model. On average, more than half of their stellar masses were formed $>$10 Gyrs ago, and more than 90\% were formed $>6$ Gyrs ago. The two types of galaxies show similarly flat profiles in a variety of stellar population parameters: old stellar ages indicated by $D4000$ (the spectral break at around 4000\AA), high stellar metallicities, large Mgb/Fe ratios indicating fast formation, and little stellar dust attenuation. In contrast, although blue spirals also formed their central regions $>$10 Gyrs ago, both their central regions and outer disks continuously form stars over a long timescale. Our results imply that, massive red spirals are likely to share some 
common processes of formation (and possibly quenching) with massive red 
ellipticals in the sense that both types were formed at $z > 2$ through 
a fast formation process.
Possible mechanisms for the formation and quenching of massive red spirals are discussed.
\end{abstract}


\section{Introduction}

One hundred years ago when Edwin Hubble pointed the 100-inch Hooker 
telescope to nearby galaxies, he discovered that the galaxies could 
be broadly divided into two distinct classes according to morphology:
spiral and elliptical. Theoretically, it is now well accepted that 
the structural and kinematic properties of a galaxy is predominantly 
determined by the aquisition and distribution of angular momentum
\citep[e.g.][]{Danovich2015}. Generally, ellipticals form by losing 
angular momentum, while spirals form by preserving and redistributing
angular momentum. In this case, an elliptical galaxy forms directly by 
efficient cooling of infalling gas and becomes red and quenched upon
its formation due to the consumption of cold gas, while a spiral galaxy 
forms in the host dark matter halo with fixed fractions of the mass and 
angular momentum of the halo, and once formed the disk grows gradually 
through continuous in-situ star formation with longstanding gas accretion 
processes \citep[e.g.][]{Fall1980,Mo1998,Dutton2007}. 

Large surveys of multiband photometry have further established that 
the morphology of galaxies is closely related to their colors, in the 
sense that spirals are usually blue and ellipticals are predominantly red 
\citep[e.g.][]{Strateva2001,Baldry2004,Bell2004,Conselice2006,Schawinski2014}. 
As blue colors are indicative of ongoing/recent star formation, this 
relation indicates that the quenching of star formation in galaxies 
may be accompanied with structural transformation. Simulations have long
shown that major majors of two spiral galaxies could end up with an 
elliptical galaxy with no/weak cold gas and star formation \citep[e.g.][]{Hopkins2006}.
On the other hand, however, major mergers of spirals could also produce 
a disk galaxy provided that the initial gas fraction of the progenitors is high 
\citep[e.g.][]{Springel2005,Robertson2006,Hopkins2009,Athanassoula2016,Sparre2017}.
More recent simulations have also suggested a non-merger
origin of both elliptical and spiral galaxies \citep[e.g.][]{Dekel2014,Zolotov2015},
where a blue ``nugget'' is firstly formed by the compaction of 
a highly disturbed disk due to violent disk instability and then 
converted to a red ``nugget'' due to a fast quenching process, 
which may or may not grow a red disk or ring-like structure by dry mergers.

The existence of a population of red (passive) spiral galaxies has 
further complicated the situation. In contrast to the monotonic color-morphology 
relation, these galaxies present spiral features but are red and quenched. 
Since the first reported case by \cite{vandenBergh1976}, many 
studies have been carried out in searching for such strange galaxies and 
also in understanding their origins\citep[e.g.][]{Dressler1999,Poggianti1999,Goto2003}. 
Thanks to large photometric and spectroscopic surveys such as the Sloan 
Digital Sky Survey (SDSS,\citealt{York2000}), our understanding of red spiral 
galaxies have significantly advanced in the past decade \citep[e.g.][]{Skibba2009,Bundy2010,
Masters2010,Robaina2012,Tojeiro2013,Fraser2018}. From these studies, 
it is generally established that red spiral galaxies are distinct from 
their blue counterparts in many aspects. Compared to blue spirals, 
basically, red spirals have more concentrated light distribution \citep{Bundy2010}
and an enhanced fraction of bars \citep{Masters2010, Fraser2018, Guo2020}, 
and they are found preferentially in environment of intermediate densities 
and dominated by LINER-like emission \citep{Masters2010}. 
Their red colors are real, thus truly reflecting the passive nature
of the galaxy, but not due to dust reddening \citep[e.g.][]{Tojeiro2013}. 

Recently, \cite{Guo2020} selected a sample of massive red spiral galaxies 
from SDSS DR7\citep{Abazajian2009} with $M_\ast > 10^{10.5}$M$_{\odot}$, 
and made comparisons with reference samples of blue spirals and red ellipticals 
of similar masses. Using SDSS single-fibre 
spectra and optical imaging data, the authors found that the red spirals are 
more similar to red ellipticals than the blue spirals in many global parameters. 
In a companion work, \cite{Hao2019} made use of the integral field spectroscopy 
from the Mapping Nearby Galaxies at Apache Point Observatory 
\citep[MaNGA;][]{Bundy2015}, and analyzed the spatially resolved stellar 
population properties for a subset of galaxies in \cite{Guo2020}. 
The similarity between red spirals and red ellipticals was confirmed in terms of global measurements of stellar population
properties including stellar age, metallicity and $\alpha$-element 
abundance, although differences are seen in stellar kinematics in 
the outer regions.

In this paper we extend the work of \cite{Hao2019} by estimating the star formation 
history (SFH) of their galaxies. Our analysis is based on the spectra fitting code, 
Bayesian Inference of Galaxy Spectra (BIGS, \citealt{Zhou2019}), which makes use 
of the full spectra fitting approach to constrain various stellar population 
properties. In our previous works, we have extensively tested the robustness of the 
code \citep{Zhou2019}, and successfully applied it to MaNGA data to constrain the 
SFHs of low-mass galaxies in the local Universe\citep{Zhou2020}.
SFHs from BIGS will provide direct evidences about the origins and formations 
of red spiral galaxies. In addition, the spatially resolved spectroscopy from 
MaNGA allows us to study the spatial variations of the SFH within individual galaxies, 
thus providing additional constraints on how the galaxies have grown. 
Moreover, MaNGA galaxies have complete photometry measurements including both 
optical photometry from SDSS and UV fluxes from the {\it Galaxy Evolution Explorer} 
(GALEX, \citealt{Martin2005}), accompanying the NASA Sloan Atlas catalogue 
\footnote{\label{foot:nsa}\url{ http://www.nsatlas.org/}}(NSA, \citealt{Blanton2005}). 
This data allows us to select truly quenched spiral galaxies, and make comparisons
with the optically selected red spirals as studied in \citet{Guo2020} and \citet{Hao2019}.

The paper is organised as follows. In \S\ref{sec:data} we present the data used 
in the analysis, including the sample selection process and a brief introduction 
to the MaNGA data used in this work. Our results are presented in \S\ref{sec:result}, 
discussed in \S\ref{sec:discussion}, and summarized in 
\S\ref{sec:summary}. A standard $\Lambda$CDM cosmology with $\Omega_{\Lambda}=0.7$, 
$\Omega_{\rm M}=0.3$ and $H_0$=70\kms Mpc$^{-1}$ is assumed throughout this work. 

\section{data and analysis}
\label{sec:data}
\subsection{MaNGA data}

\begin{figure*}
    \includegraphics[width=0.5\textwidth]{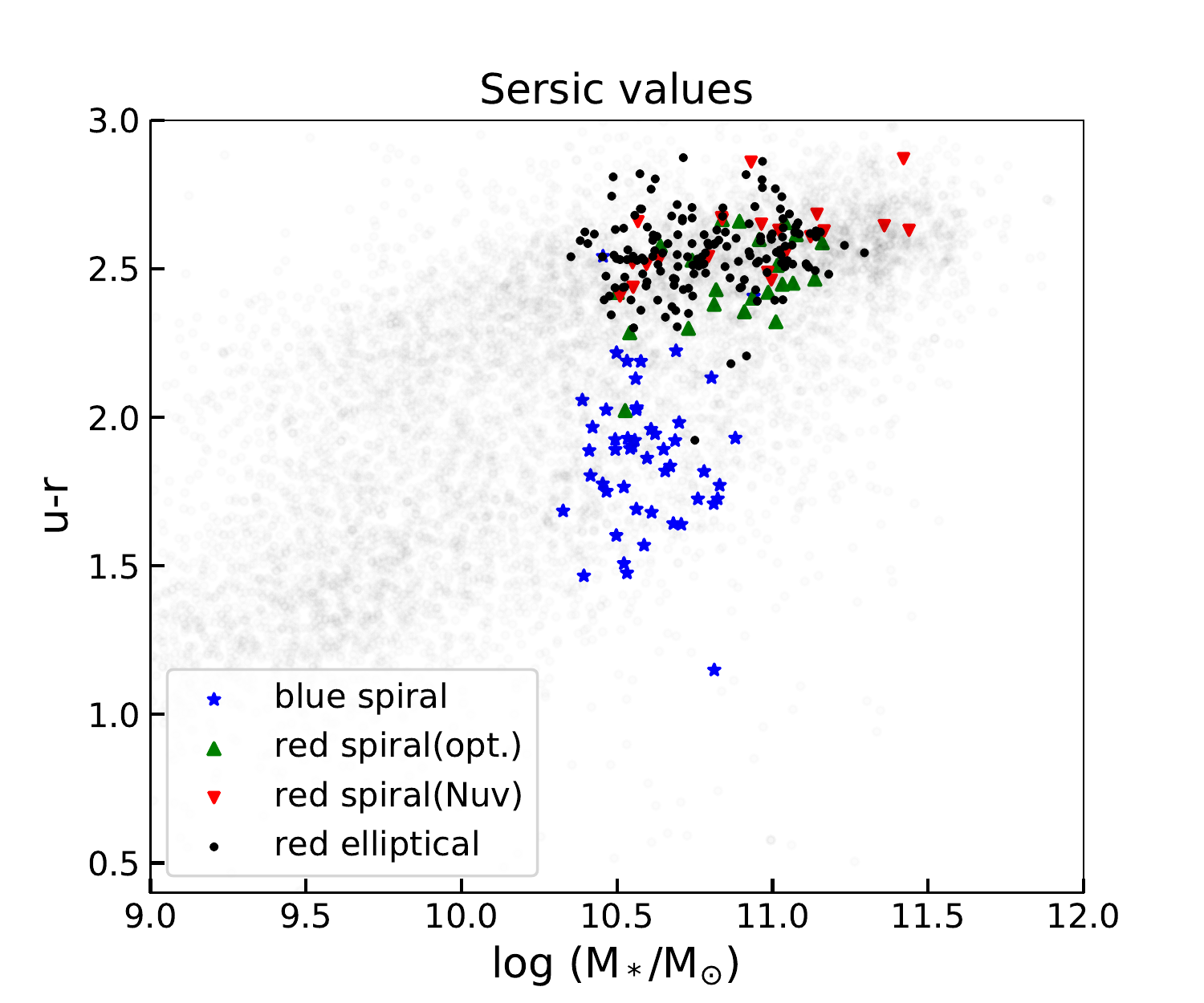}
    \includegraphics[width=0.5\textwidth]{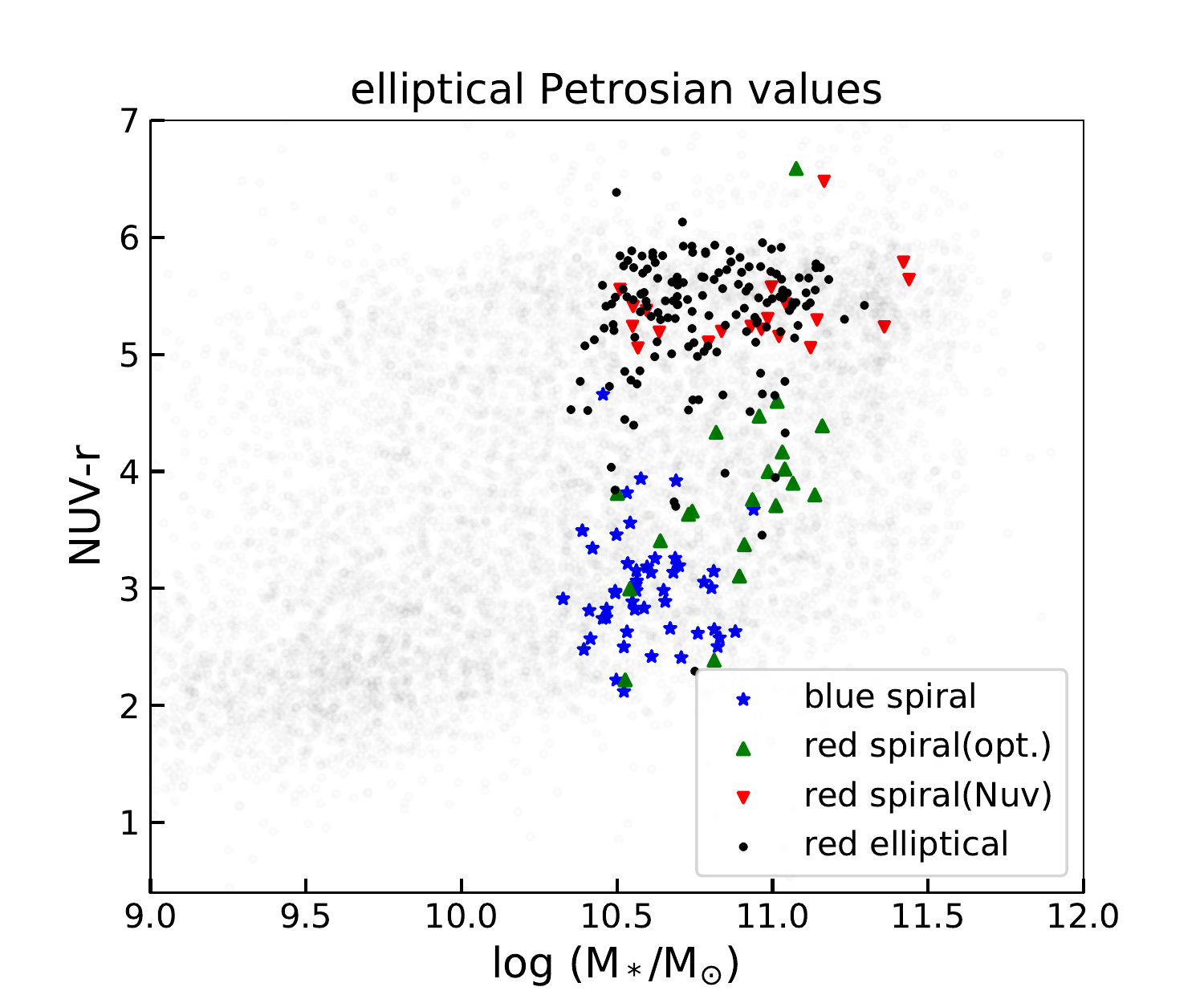}
    \caption{Our samples on the $u-r$ vs. M$_*$ (left) and $NUV-r$ vs. 
    M$_*$ (right) planes. The $u-r$ colors in the left panel are based 
    on model magnitudes from a single S\'{e}rsic profile, while 
    the $NUV-r$ colors in the right panel are based on the elliptical 
    Petrosian model. Galaxies of red spirals(opt.), red spirals(NUV), 
    blue spirals and red ellipticals are plotted with different colors 
    and symbols, as indicated. Plotted in background grey dots are 
    the whole MaNGA DR15 sample.}
    \label{fig:sample}
\end{figure*}

MaNGA is one of the three core programs of Sloan Digital Sky Survey IV 
\citep[SDSS-IV][]{Blanton2017}), aiming at obtaining spatially resolved spectroscopy 
for $\sim$10,000 galaxies in the low-$z$ Universe ($0.01 < z < 0.15$) . 
MaNGA targets are selected from the SDSS with stellar mass in the range 
$5\times10^8 {\rm M}_{\odot}h^{-2} \leq M_*\leq 3 \times 10^{11} {\rm M}_{\odot}h^{-2}$ 
\citep{Yana2016,Wake2017}. Targets in the “Primary” and “Secondary” samples 
are each covered out to either 1.5 or 2.5 effective radius($R_e$) \citep{Law2015}. 
The targets are observed with the Sloan 2.5 m telescope \citep{Gunn2006} 
with the two dual-channel BOSS spectrographs\citep{smee2013}, which 
provide high resolution ($R\sim2000$, \citealt{Drory2015}) spectra in the 
wavelength range $3600-10300$ {\AA}. Raw data of MaNGA are reduced with 
the official data reduction pipeline(DRP, \citealt{Law2016}) to produce 
sky-subtracted, spectrophotometrically calibrated spectra that are ready 
for scientific studies. Flux calibrations of MaNGA spectra are better than  
$5\%$ for most of the wavelength ranges \citep{Yana2016,Yanb2016}. 
In addition, data products including stellar kinematics, emission line properties 
and spectral indices are provided by the MaNGA Data Analysis Pipeline 
\citep[DAP,][]{Westfall2019,Belfiore2019}. The MaNGA DRP and DAP 
data have been released in SDSS/DR15 \citep{Aguado2019}
for 4,621 unique galaxies.

\subsection{Sample selection}
\label{ssec:sampleselection}

In a recent work, \cite{Guo2020} has constructed a sample of massive 
red spirals, as well as samples of massive ellipticals and blue spirals,
from the SDSS main galaxy sample. We take their samples and match 
each sample with the sample of MaNGA. This gives rise to three samples of massive galaxies ($M_\ast>10^{10.5}{\rm M}_{\odot}$), respectively including 22 red spirals, 49 blue spirals and 158 red ellipticals. 
The reader is referred to \citet{Guo2020} for details about the 
selection criteria of the different types. In short, massive galaxies
with stellar masses larger than $10^{10.5} {\rm M}_{\odot}$ 
from the SDSS DR7 \citep{Abazajian2009} are selected in the first place,
and are divided into spirals and elliticals according to morphological 
classifications taken from the Galaxy Zoo I project \citep{Lintott2011}. 
Each galaxy in samples of both spirals and ellipticals was then 
classified to be blue or red according to optical $u-r$ color
(corrected for dust attenuations), for which model magnitudes estimated 
from a single S\'{e}rsic profile were adopted \citep{Blanton2011}. 
During this selection process, galaxies either with minor-to-major axis 
ratio $b/a<0.5$ or visually inspected to be edge-on are excluded so as 
to minimize the effect of dust attenuation and to ensure correct 
disk-bulge decomposition. 

The left panel of \autoref{fig:sample} displays the galaxies of 
the three classes in the plane of $u-r$ (S\'{e}rsic model) versus 
$\log_{10}M_\ast$. To keep consistency with MaNGA which selected 
targets from NSA, we use model magnitudes and stellar masses also
from the NSA. When compared to the official SDSS photometric catalogs, 
the NSA is expected to provide better measurements for sizes and 
luminosities of nearby large galaxies, as the SDSS imaging were 
reprocessed with substantial improvements in both background subtraction 
and de-blending \citep{Blanton2011}. 
The colors and stellar masses from NSA are slightly different 
from those used in \cite{Guo2020}, where the colors were taken 
from the SDSS photometric catalogs and the masses were from
\cite{Mendel2014}. This is why the stellar 
mass of some galaxies fall slightly below the mass limit of
$10^{10.5}M_\odot$, and a few red spirals/ellipticals fall in the
region of blue spirals. For the majority of the galaxies, however,
the color classification is consistent with \citet{Guo2020}. 

When visually inspecting the optical image of the red spirals, 
we found that many of them have quite significant color gradients: 
the outer regions generally look bluer, which hints that those 
galaxies may not be totally quenched. This echoes the early work 
by \citet{Cortese2012} who found optically-selected passive spirals 
in \citet{Masters2010} presented significant star formation and 
$NUV$ emission. In the right-hand panel of \autoref{fig:sample}, 
we plot the three samples of galaxies 
in the plane of $NUV-r$ color versus $\log_{10}M_\ast$. It is known 
that the $NUV-r$ color is more sensitive to the residual cold 
gas and weak star formation in galaxies, when compared to optical 
colors. It is seen that the blue spirals are mostly blue 
with $NUV-r<4$, but some of the red ellipticals fall in the 
``green-valley'' regime with $4<NUV-r<5$. In agreement with 
\citet{Cortese2012}, all the optically-selected red spirals in our 
sample are actually green or even blue with $NUV-r<5$ in this diagram. 
The comparison between the $u-r$ and $NUV-r$ colors highlights the 
necessity of using near-ultraviolet instead of optical colors for 
selecting fully quenched galaxies. We thus select an additional sample of massive 
red spirals on the $NUV-r$ versus $\log_{10}M_\ast$ diagram.  
We make use of the morphology classification from 
\cite{Dominguez2018}, in which the morphology type of a galaxy 
is indicted by its $T$-type number: with T-type$\leq$0 for early-type 
galaxies (ETGs), T-type$>$0 for late-type galaxies (LTGs). 
We also exclude edge-on galaxies by both using $b/a$ and visually 
examining the SDSS images. We obtain a sample of 20 
$NUV$-selected red spiral galaxies with $T$-type$>$0, 
$NUV-r>5$ and M$_*>10^{10.5} {\rm M}_{\odot}$. 
In what follows, this sample is referred as ``red spirals (NUV)'', 
and the 22 red spirals selected above from the $u-r$ color 
is referred as ``red spirals (opt.)''. 

Again, to keep consistency with MaNGA, we have adopted 
elliptical Petrosian magnitudes instead of S\'{e}rsic 
magnitudes to define the $NUV-r$ color, as MaNGA uses 
elliptical Petrosian measurements of effective radius (Re) 
and flux for sample selection (see \citealt{Wake2017} 
for details). We have examined our samples on the $NUV-r$ 
versus mass diagram using S\'{e}rsic magnitudes, finding 
the optically-selected red spirals to be slightly redder 
but still falling in the green-valley area. Locations of 
other samples of galaxies including the $NUV$-selected red 
spirals remain unchanged. The redder $u-r$ with S\'{e}rsic 
magnitudes can be understood as a combined effect of two 
facts: on one hand the S\'{e}rsic model covers an effective 
radius ($R_e$) that is smaller than the Petrosian model, 
and on the other hand galaxies typically present a negative 
color gradient with redder colors at smaller radii. 
In addition, we have examined the morphology type from 
Galaxy Zoo for the $NUV$-selected red spirals, and confirmed 
that they indeed present spiral feature in their SDSS image, 
though weak in most cases. Therefore, our results to be 
presented in the next section are not affected by the choice 
of magnitude definition or morphology classification method. 
\begin{figure*}
    \centering
    \includegraphics[width=0.32\textwidth]{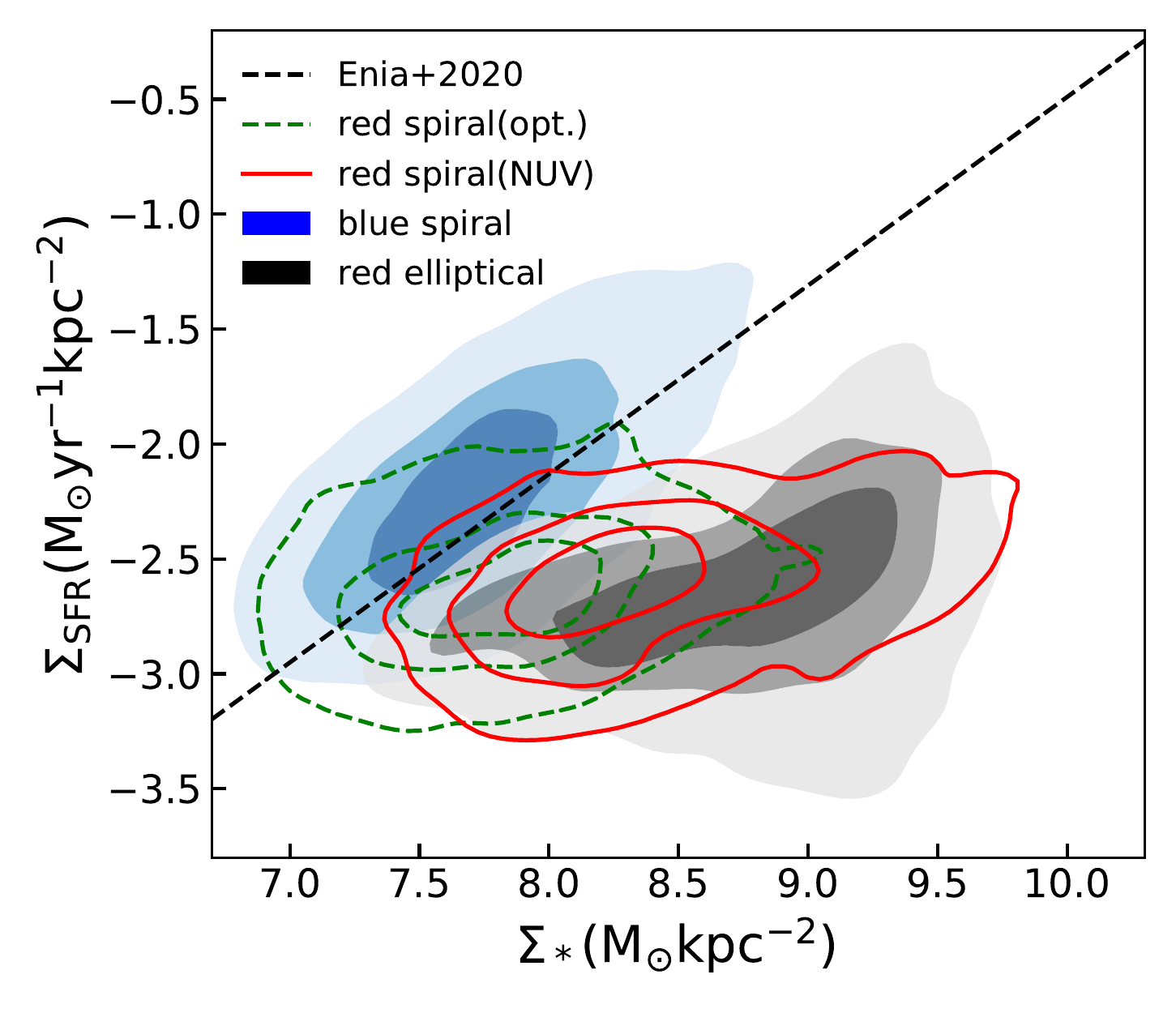}
    \includegraphics[width=0.32\textwidth]{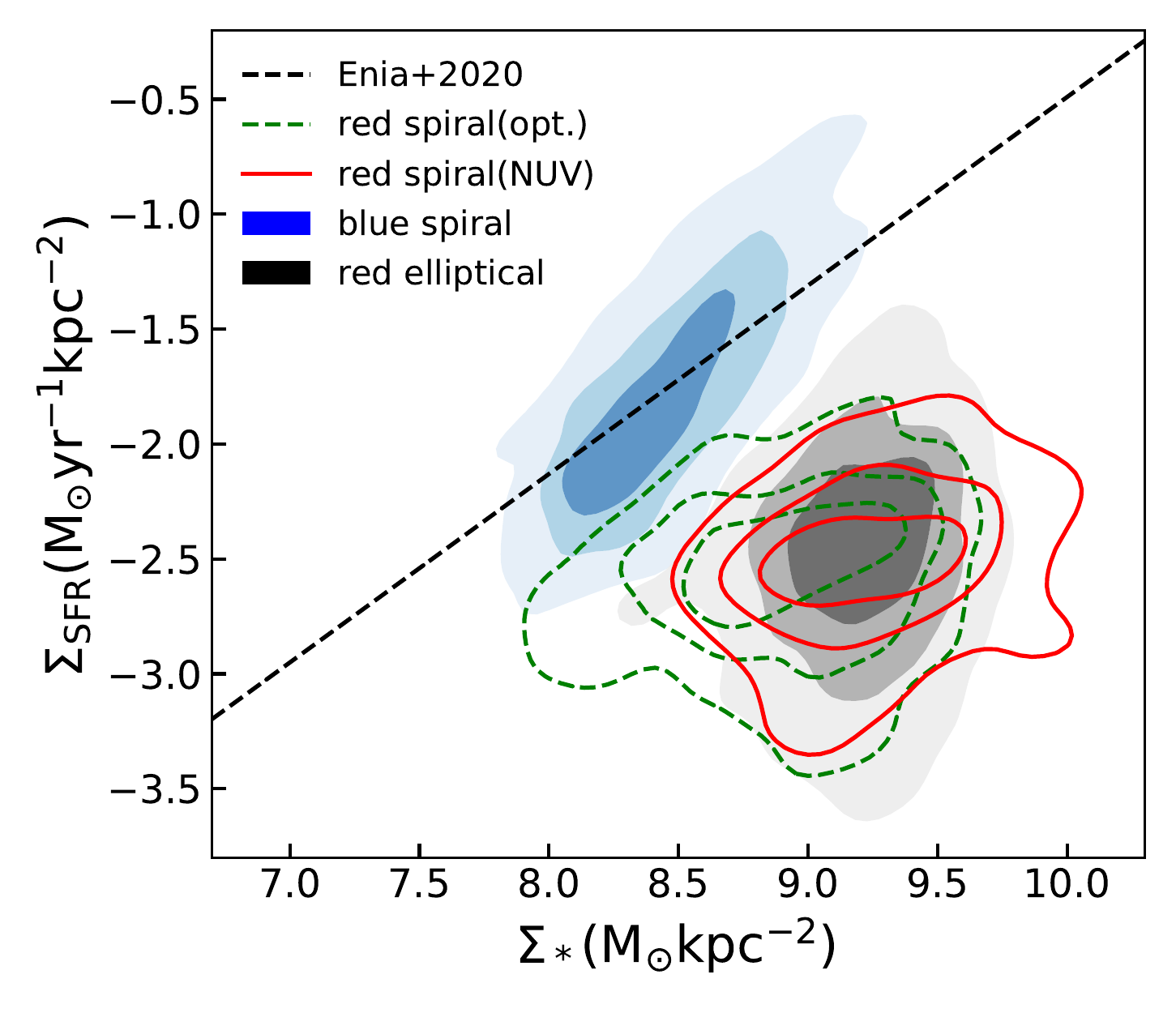}
    \includegraphics[width=0.32\textwidth]{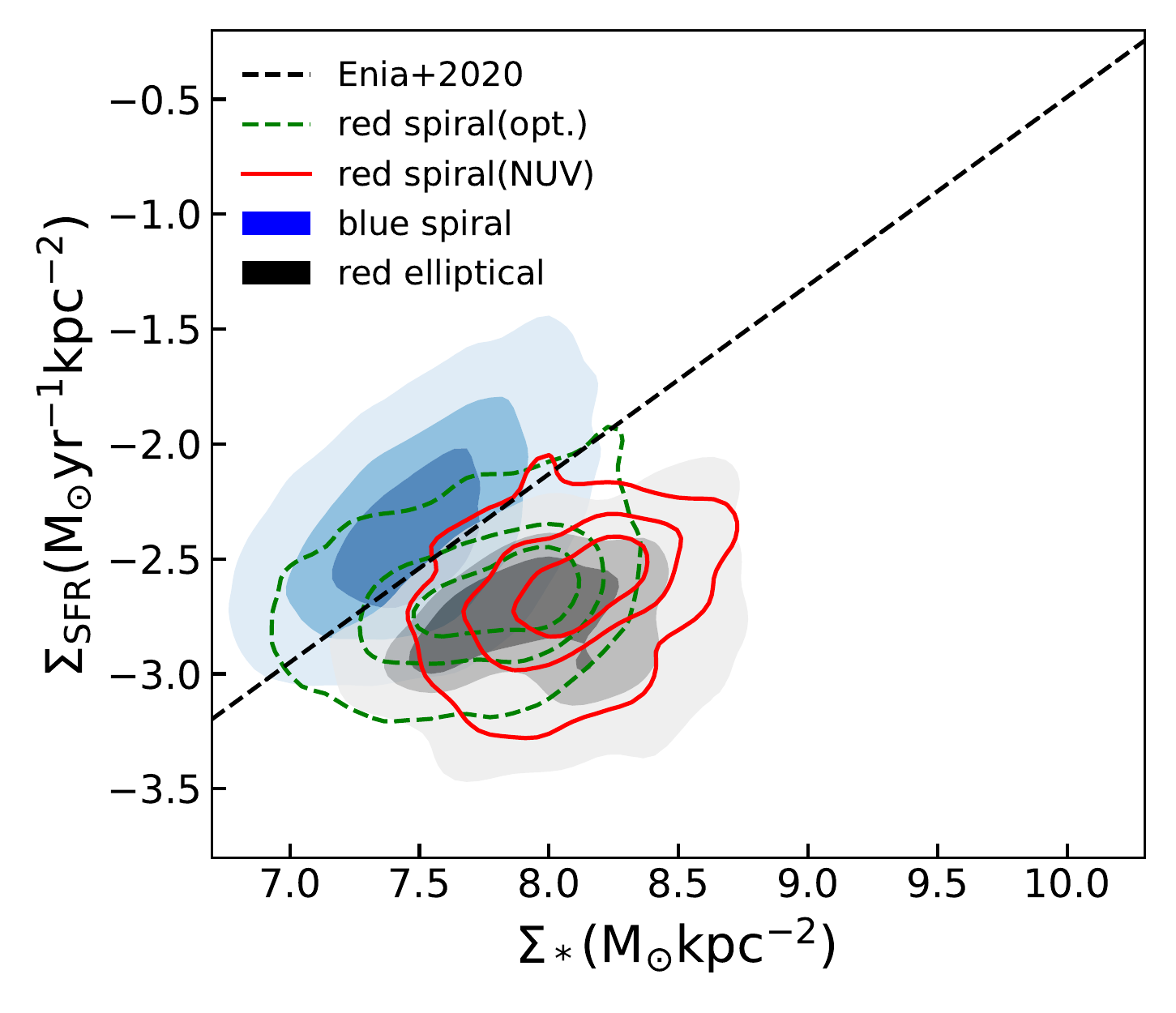}
    \caption{ The $\Sigma_{\rm SFR}$-$\Sigma_*$ relations for spaxels of blue spiral(blue), red spiral(opt., green), red spiral(NUV, red) and red elliptical(grey) galaxies. The left panel shows contours for all spaxels, while the middle and right panels show spaxels from the central($<$0.5Re) and outer($>$1.0Re) regions. The contours hold 30\%, 60\% and 90\% of the total amount of data points. The black dash line shows the main sequence from \cite{Enia2020} as reference.}
    \label{fig:SFMS}
\end{figure*}

\autoref{fig:SFMS} shows the star formation rate surface 
density ($\Sigma_{\text{SFR}}$) against stellar mass surface density 
($\Sigma_\ast$) for the four samples. For this, we use H$\alpha$ flux 
measurements from MaNGA DAP and a conversion factor of: 
$\rm SFR(M_{\odot}yr^{-1})=7.9\times10^{-42}L(H\alpha)$ from 
\cite{Kennicutt1998ARA} to estimate the current star formation rate 
of a given spaxel in our sample galaxies. The stellar mass measurements
are estimated from the MaNGA spectra by \cite{Li2020}, by applying 
a stellar population synthesis code with BC03 stellar population 
models and assuming a Chabrier IMF. Results of all spxels from the
different samples are shown in the left panel of the figure. As expected,
the blue spiral galaxies well follow the resolved star-forming main 
sequence, indicated by the black dashed line from \cite{Enia2020}, 
while the red ellipticals mostly fall below this sequence. 
A non-negligible fraction of spaxels from optically 
selected red spirals are still forming stars, while the $NUV$-selected 
red spirals are almost quenched entirely, occupying similar regions 
to the red ellipticals. These trends are well 
consistent with the global $NUV-r$ color as seen in the previous
figure. It is remarkable that the star forming spaxels from red 
spirals (Opt.) are mostly from low surface mass density regions. 
When examining the central ($<$0.5$R_e$, the middle panel) and 
outer ($>$1.0$R_e$, the right panel) regions separately, 
we see that the low surface mass density regions are mostly located 
in the outer region, while the central high density regions of all 
types of red spirals are almost quenched. In what follows, we will 
investigate the underlying stellar populations and their formation
histories for the different types of galaxies in more detail.

\subsection{Spectral analysis}

\begin{figure*}
    \centering
    \includegraphics[width=0.8\textwidth]{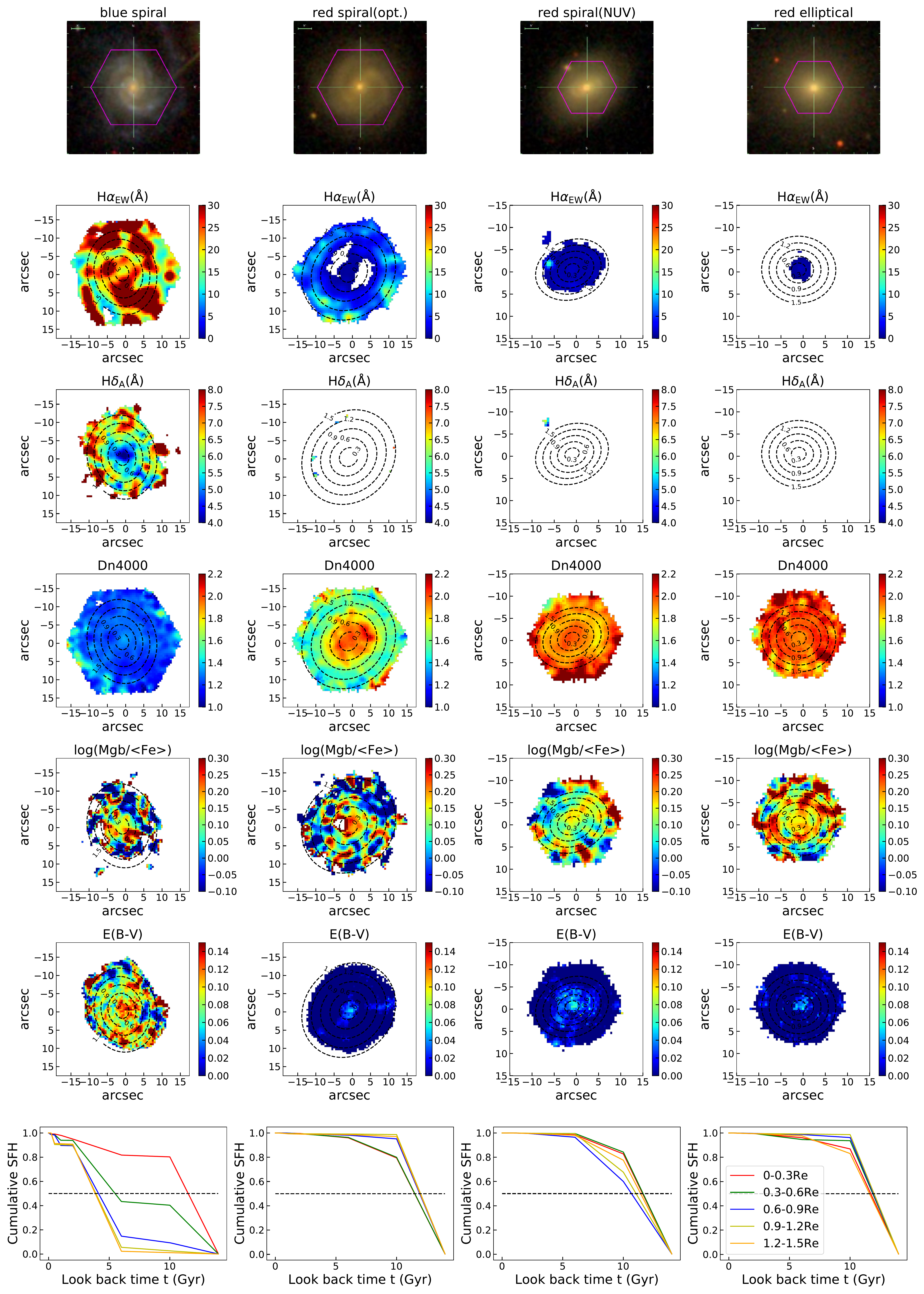}
    \caption{Four columns each shows an example galaxy in the four samples used in this work respectively as labelled. In each column,
The first panel is the optical image with MaNGA footprint showing in
magenta. The second to 5th panels show maps of H$\alpha$ equivalent width, H$\delta_{\rm A}$, Dn4000, and log(\mgbfe) respectively from  MaNGA DAP. The 6th panel plots the E(B-V) map from \cite{Li2020}. Dash lines in each map denotes the different Re bins used in the stacking processes. The last panel shows the cumulative SFH of different radius bins in this galaxy, derived from the best-fit models of stepwise SFH(see \S \ref{ssec:sfh}). 
    }
    \label{fig:example}
\end{figure*}

We estimate the stellar population properties of galaxies
in our samples using {\tt BIGS}, Bayesian Inference of galaxy 
spectra, which is a Python spectral fitting code developed in our previous 
paper \citet{Zhou2019}. This code has been applied to MaNGA data 
to constrain stellar initial mass function (IMF) in elliptical 
galaxies \citep{Zhou2019} and star formation history (SFH) of 
low-mass galaxies \citep{Zhou2020}. Here we briefly describe the 
process of applying this code to the samples of massive galaxies 
as selected above. We refer the reader to our previous papers 
for a more detailed description of the fitting algorithm and tests. 

For a given galaxy, we take all the spaxels within 1.5$R_e$ and 
divide them into five radial bins with a fixed width of 0.3$R_e$, 
according to their elliptical annuli radii given by MaNGA DAP. 
Spectra in a common radial bin are then stacked following the 
approach in \cite{Zhou2020} to achieve enough signal-to-noise ratio
(SNR) required by subsequent analysis. The correction term given 
by \cite{Westfall2019} is used to account for the covariance between 
spaxels. In our sample, the SNR (averaged over all wavelengths
for a given spectrum) of the stacked spectra is typically above 
100 per {\AA} in the innermost radial bin, and decreases to around 
40 per {\AA} in the outermost bin. In addition, the SNR 
depends on wavelength in the sense that the SNR is highest at 
around 6500{\AA}, decreasing slightly towards both shorter and
longer wavelengths. Each of the stacked spectra is firstly fitted 
using a simple spectral fitting code developed in \cite{Li2020}, 
giving rise to an estimate of the stellar velocity dispersion 
($\sigma_\ast$) that accounts for effects of both stellar kinematics 
and instrumental resolution. Emission lines in the spectrum are 
identified in this procedure, and are masked out in the following 
stellar population synthesis modelling. The wavelength range 
6800-8100 {\AA} with known mismatchments between template and 
observed continua is also masked out following \citep{Zhou2020}. 

We then model the stellar component of each stacked spectrum with 
the following stellar population synthesis approach. 
To begin with, the stellar initial mass function of \cite{Chabrier2003}
and a specific star formation history (SFH, see below) are adopted. 
The SFH model is combined with the E-MILES \citep{Vazdekis2016} 
SSP templates and a simple screen dust model \citep{Charlot2000} 
to generated composited model spectra, which are then convolved with the 
stellar velocity dispersions derived above to account for kinematic 
and instrumental broadening effects. The model spectra 
are then compared with the observed spectrum to calculate the 
likelihood for a given set of model parameters ($\theta$):
\begin{equation}
\label{likelyhood}
\ln {L(\theta)}\propto-\frac{1}{2}\sum_{i,j=1}^N\left(f_{\theta,i}-f_{D,i}\right)\left({\cal
M}^{-1}\right)_{ij}\left(f_{\theta,j}-f_{D,j}\right)\,
\end{equation}
where $f_{\theta, i}$ is the flux at the $i$-th wavelength point 
as predicted for the parameter set $\theta$, $f_{D, i}$ is the 
flux at the same wavelength in the stacked spectrum, and $N$ the 
total number of wavelength points. The matrix ${\cal M}_{ij}\equiv
\langle \delta f_{D, i}\delta f_{D, j} \rangle$
is the covariance matrix of the data, which in this case is 
diagonal and specified by the error vectors produced in the 
stacking procedure. {\tt BIGS} utilizes the {\tt MULTINEST} 
sampler \citep{Feroz2009,Feroz2013} and its \textsc{Python} interface\citep{Buchner2014} to sample the posterior distributions 
of the model parameters and derive the Bayesian evidence. 

The purpose of this study is to reveal the SFH of the different 
types of massive galaxies. To the end, we consider three different 
models to characterise the SFH of our galaxies: the $\Gamma$ model, 
the $\Gamma$+B model and the stepwise model, are used to characterise 
the SFHs of the galaxies. The $\Gamma$ model simply uses a $\Gamma$ 
function to describe the star formation rate as a function of cosmic time:
\begin{equation}
    \Psi(t)=\frac{1}{\tau\gamma(\alpha,t_0/\tau)} 
\left({t_0-t\over \tau}\right)^{\alpha-1}
    e^{-(t_0-t)/\tau}\,,
\label{gamma-sfh}
\end{equation}
where $t_0$ is the present-day time, $t$ is the look back time, and 
$\gamma(\alpha,t_0/\tau)\equiv \int_0^{t_0/\tau} x^{\alpha-1}e^{-x}\,dx$.
The $\Gamma$+B model is a $\Gamma$ model plus an additional burst 
characterised by a SSP, where the time of the burst is a free parameter. 
The stepwise model describes the SFH in a 
non-parametric way, using the average star formation rates in 7 time 
intervals over the whole history of star formation. Specifically we 
adopt the following time intervals: $0\to 0.2$, $0.2\to 0.5$, 
$0.5\to 1.0$, $1\to 2$, $2\to 6$, $6\to 10$ and $10 \to 14$ Gyr. 

In addition to the spectral fitting approach, we also make use 
of absorption and emission line features that are sensitive to 
star formation. We use $Dn4000$ (the spectral break at 4000\AA), 
$EW$(H${\rm \delta_A}$) (equivalent width of the H$\delta$ absorption line), 
\mgbfe $\rm \equiv Mgb/(0.5*Fe5270+0.5*Fe5335)$ (the Mg-to-iron 
abundance ratio), and $EW$(H$\alpha$) (equivalent width of the 
H$\alpha$ emission line) derived by the MaNGA DAP. Moreover, we
have derived stellar dust attenuation maps as quantified by $E(B-V)$ 
using the method developed by \cite{Li2020} for our galaxies.
These parameters provide independent and complementary probes of 
recent SFHs, in addition to the SFH constrains derived from the 
spectral fitting. 

\autoref{fig:example} shows the stellar parameters and the SFH 
for four typical galaxies in our sample, each selected from one 
of the four samples. Their SDSS images are shown in the top row, 
with the hexagonal IFU of MaNGA indicated. 
The blue spiral galaxy in the left-most column shows 
high values of $EW$(H$\alpha$) and $E(B-V)$ and low values of 
$Dn(4000)$ with relatively weak radial gradients, indicative of 
ongoing star formation and young populations across the whole galaxy. 
In contrast, the elliptical galaxy in the rightmost column 
shows no/weak H$\alpha$ emission and H$\delta$ absorption, 
$Dn(4000)>1.6$ and small $E(B-V)$ everywhere, indicating that 
the entire galaxy is fully quenched and dominated by old populations.  
The $NUV$-selected red spiral (the third column) looks similar 
to the elliptical in all the parameters. The optically-selected 
red spiral (the second column) is similar to the blue spiral
in the outskirts, while its inner region look similar to the 
elliptical and $NUV$-selected red spiral. This is consistent 
with the $NUV-r$ versus mass diagram where the optically-selected 
red spirals fall in between the red and blue populations 
(see \autoref{fig:sample}). The bottom panels show the cumulative
SFHs constrained with {\tt BIGS} for the stepwise model. It is 
interesting to see that both the optically-selected and $NUV$-selected 
red spirals present quite similar SFHs to the elliptical galaxy, 
in which a substantial fraction of the present-day stellar mass 
in the galaxies were formed at early times. In contrast, the 
blue spiral galaxy shows early formations only in the innermost 
region ($R<0.3R_e$), and the outer regions show much extended SFHs. 
In the following section we will present the statistical behaviors
of the SFHs and stellar population parameters for our samples.

\section{Results}
\label{sec:result}

\subsection{Star formation histories}
\label{ssec:sfh}

\begin{figure*}
    \centering
    \includegraphics[width=1.0\textwidth]{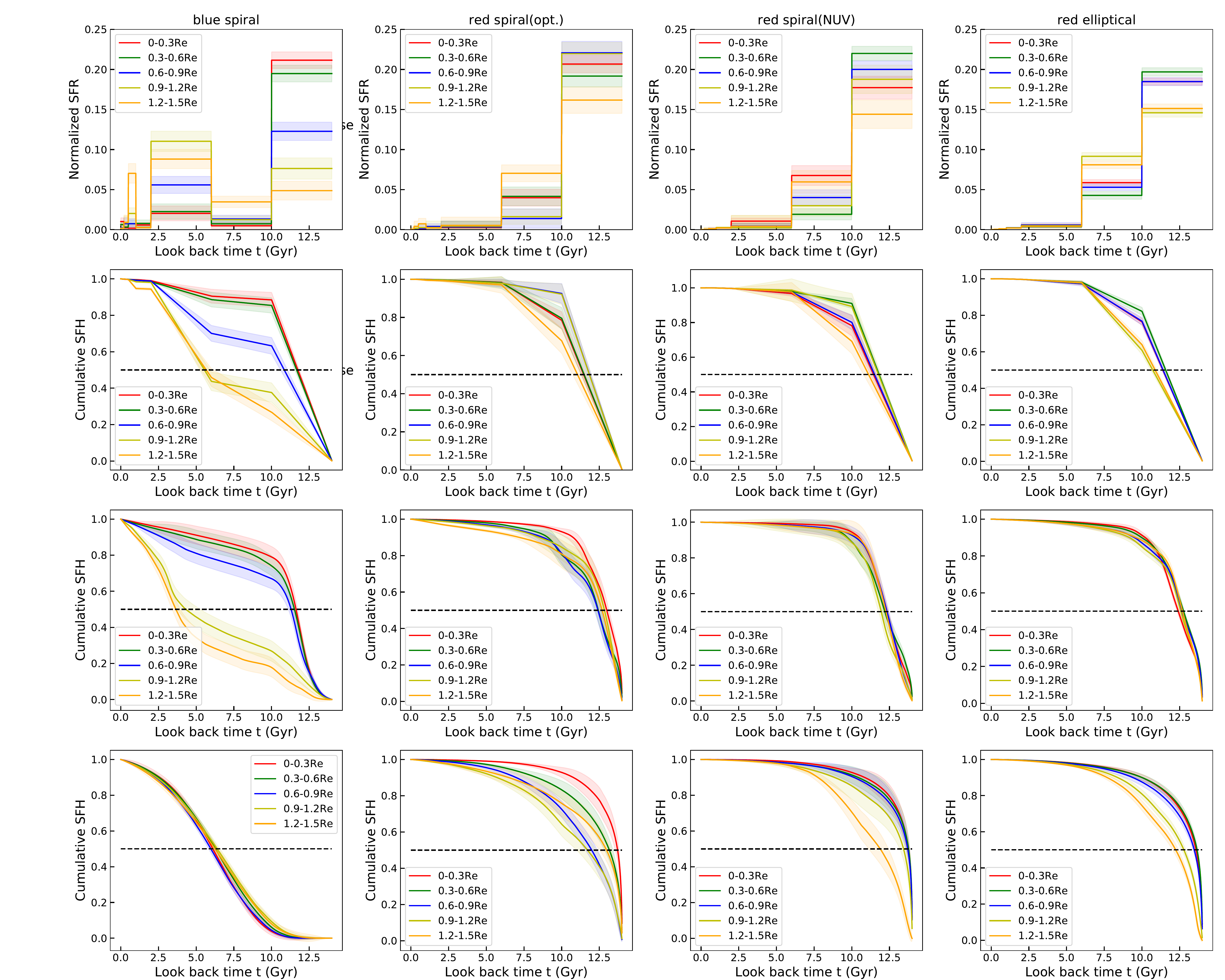}
     \caption{Star formation histories from the best-fit models. The four columns are results of the four samples respectively, as labelled. The first row plots SFHs from the stepwise model, with the second row showing the corresponding cumulative SFHs. As comparisons, the third and fourth rows are cumulative SFHs from the $\Gamma$+B model and $\Gamma$ model, respectively. In each panel, lines are medians of the SFH over the sample galaxies, with shaded regions showing the uncertainties. Results from the five radius bins are marked with different colors, as labelled.}
     \label{fig:sfh_nre}
\end{figure*}

\begin{figure*}
    \centering
    \includegraphics[width=0.45\textwidth]{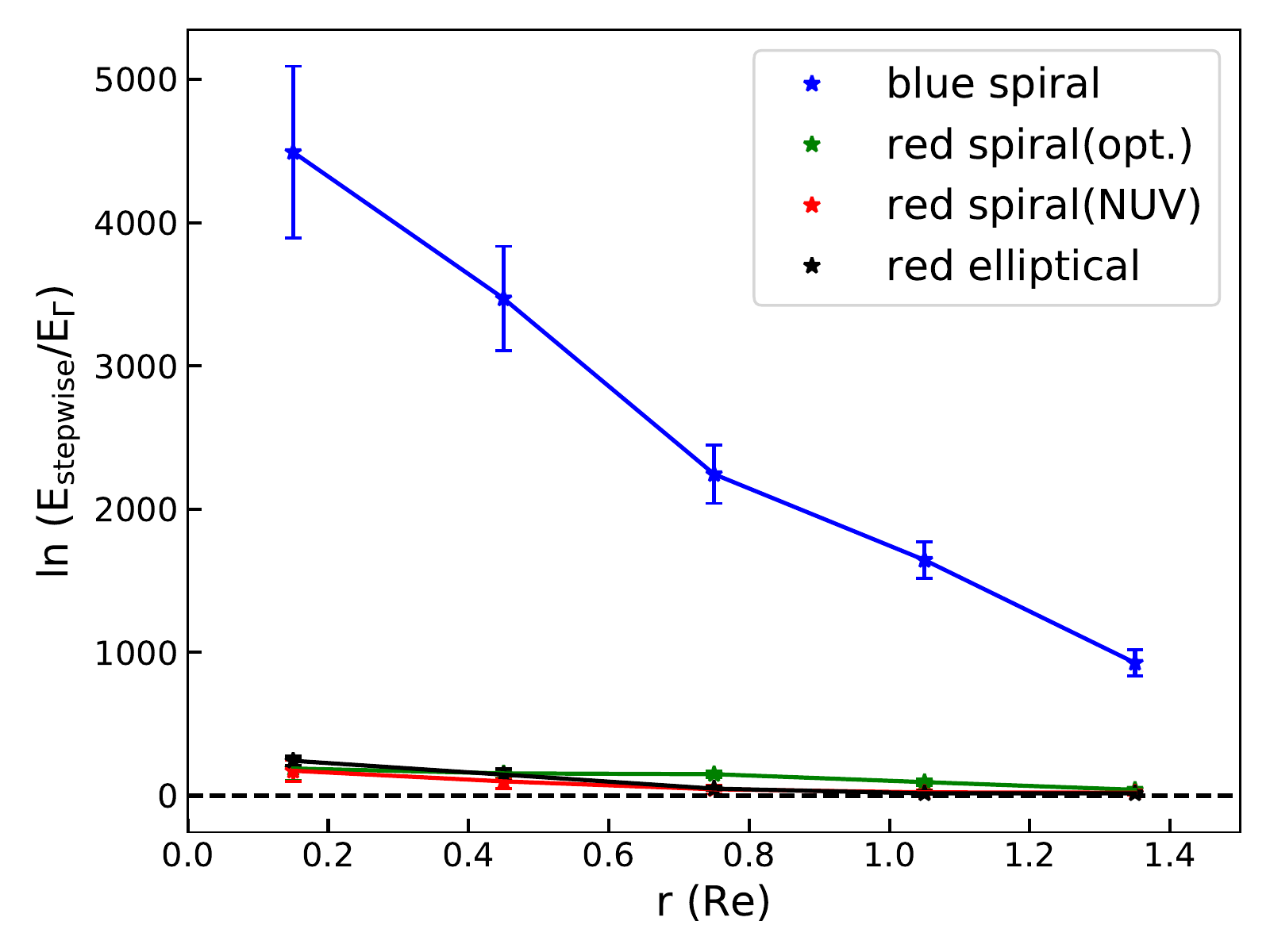}
    \includegraphics[width=0.45\textwidth]{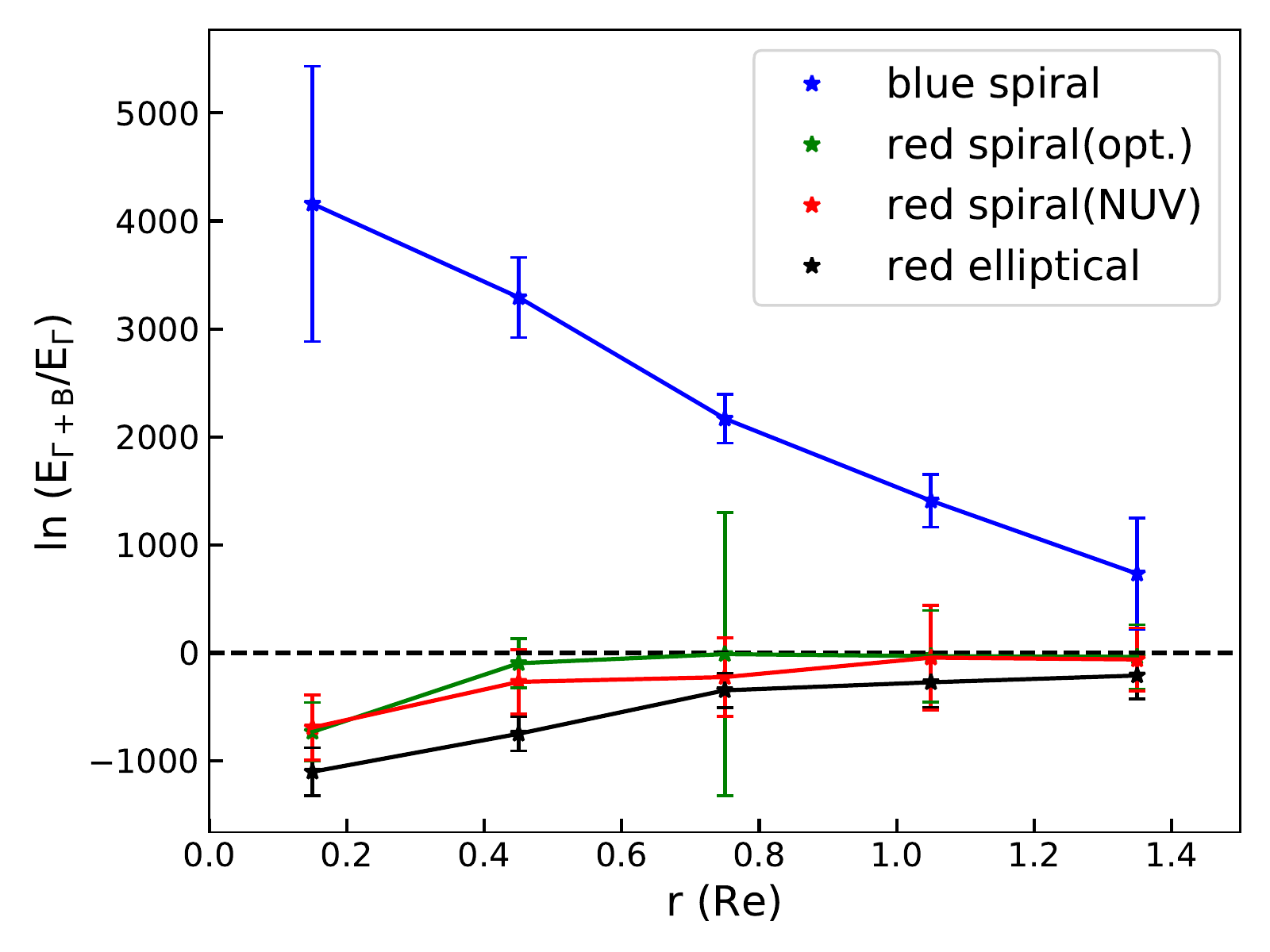}
    \caption{The evidence ratio between the stepwise and 
$\Gamma$ models (left) and between the $\Gamma$+B and 
$\Gamma$ models(right) as a function of radius. Green, red, black and blue lines show the median values of blue spiral, red spiral(opt.), red spiral(NUV), and red elliptical respectively. Error bars are obtained from the jackknife resampling method.
    }
    \label{fig:modelselection}
\end{figure*}

We obtain the best-fit SFH for different radial bins of each galaxy 
in our sample from the posterior distribution of the model parameters,
produced by {\tt BIGS} as described above. The results are shown in 
\autoref{fig:sfh_nre}, where we show the median SFH in each of the 
five radial bins for the four types of galaxies (panels from left to 
right): blue spirals, red spirals (opt.), red spirals (NUV) and 
red ellipticals. For each type, we show in the top panel the 
differential SFH in seven time intervals from the stepwise model, 
and the cumulative SFHs from the stepwise model, the $\Gamma$+B
model and the single $\Gamma$ model in the lower panels. 
It is seen that for red ellipticals and red spirals (both optically- 
and $NUV$-selected), the majority of the current stellar mass
were formed in the oldest time bin ($>10$Gyr), which contributed 
a fraction of mass as high as 80-90\% at the smallest radius and 
$\ga 60$\% in the outskirt of galaxies. The intermediate time 
interval (6-10Gyr) contributes the rest $\sim10$\% in the smallest 
radial bin and the rest $\sim30-40$\% in the outskirt. The 
three types of galaxies have been almost fully quenched in the past 
6Gyr, contributing at most only a few percent of the total stellar mass. 
The results appear to be hold regardless of SFH models. The stepwise 
and $\Gamma$+B models give quite consistent results, while 
the single $\Gamma$ model produces similar results in the inner 
regions but requires slightly extended SFHs in the outer regions of the 
galaxies. When compared to $NUV$-selected red spirals, optically-selected 
red spirals show subtle residual star formation in the recent 2Gyr 
in the outermost bin(1.2-1.5Re, see the top panels of the figure). 
This result is consistent with the bluer colors and weak H$\alpha$ 
emissions as seen in the outer regions of their SDSS image and 
MaNGA datacubes (mostly in their spiral arms). Overall, one can 
conclude that both red spirals and ellipticals are formed at 
early times ($z\ga 2$), with little star formation afterwards. 

In contrast to red spirals and ellipticals, blue spirals show 
different SFHs at different radii in both stepwise and $\Gamma$+B 
models, in the sense that the inner regions formed stars at earlier 
times and the outer regions formed stars with multiple events 
over a longer timescale. It is interesting that, although the 
outer regions of blue spirals experience extended SFH, their 
central regions ($<0.6R_e$) behave similarly to red ellipticals 
and red spirals, where more than 80\% of the stellar mass was 
accumulated more than 10 Gyrs ago. When a single $\Gamma$ 
model is adopted, the blue spirals appear to have formed their 
stars with a quite extended history that is independent of 
radius, with half of the total mass formed at $\sim6$Gyr, 
$\sim10$\% in the recent 2Gyr and nearly no stars older than 10Gyr.

In the Bayesian context, the Bayesian evidence ratio represents 
the posterior probability for two competing model families, and 
can thus serve as a discriminator. \autoref{fig:modelselection} 
shows the evidence ratio between the stepwise and $\Gamma$ models 
(left) and the ratio between the $\Gamma$+B and $\Gamma$ models (right), 
as function of radius. The median values of the galaxies in a 
given radial bin are plotted to represent the global trend, 
with error bars estimated from the jackknife resampling method . 
It is found that, for red ellipticals and both optically-selected 
and $NUV$-selected red spirals, different SFH models give rise 
to comparable Bayesian evidence, indicating that the inferred 
SFH of the three types of galaxies are robust to the assumed 
functional form of SFH. Blue spirals show large evidence ratios 
in both panels, with larger ratios at smaller radii. This strongly 
suggests that a single $\Gamma$ model is unacceptable in order 
to describe the SFH of blue spirals, and the $\Gamma$+B or 
stepwise SFH model is preferred by the data. 

The three SFH models, $\Gamma$, $\Gamma$+B and stepwise, 
used in our analysis have different levels of flexibility: 
$\Gamma$ model can describe only one major star formation event, 
$\Gamma$+B would allow an additional burst happened in the SFH, 
while the stepwise model can catch multiple star formation events 
although with poor time resolutions. The model select results 
clearly confirm that, similar to red ellipticals, the red spirals 
are very likely to have only one major star formation episode, and 
can be characterised by a very simple SFH model, while the star 
formation histories of blue spirals can be quite complicated and 
more flexible SFH models are needed to fit the data. 

In summary, the analysis of SFHs suggest that massive red galaxies,  
both spiral and elliptical, are mostly formed in one major star 
formation episode at early universe (10Gyr ago). This result is robust 
against the assumed SFH models. Massive blue spirals 
have experienced different SFHs, with more than one major star 
formation episodes over a long timescale, and their SFHs cannot 
be correctly probed with a simple model such as the $\Gamma$ model. 
In addition, we have seen radial variations of the SFHs in 
our galaxies, which is discussed below in more detail.  

\subsection{Radial gradients}

\label{subsec:stat}
\begin{figure*}
    \centering
    \includegraphics[width=1.0\textwidth]{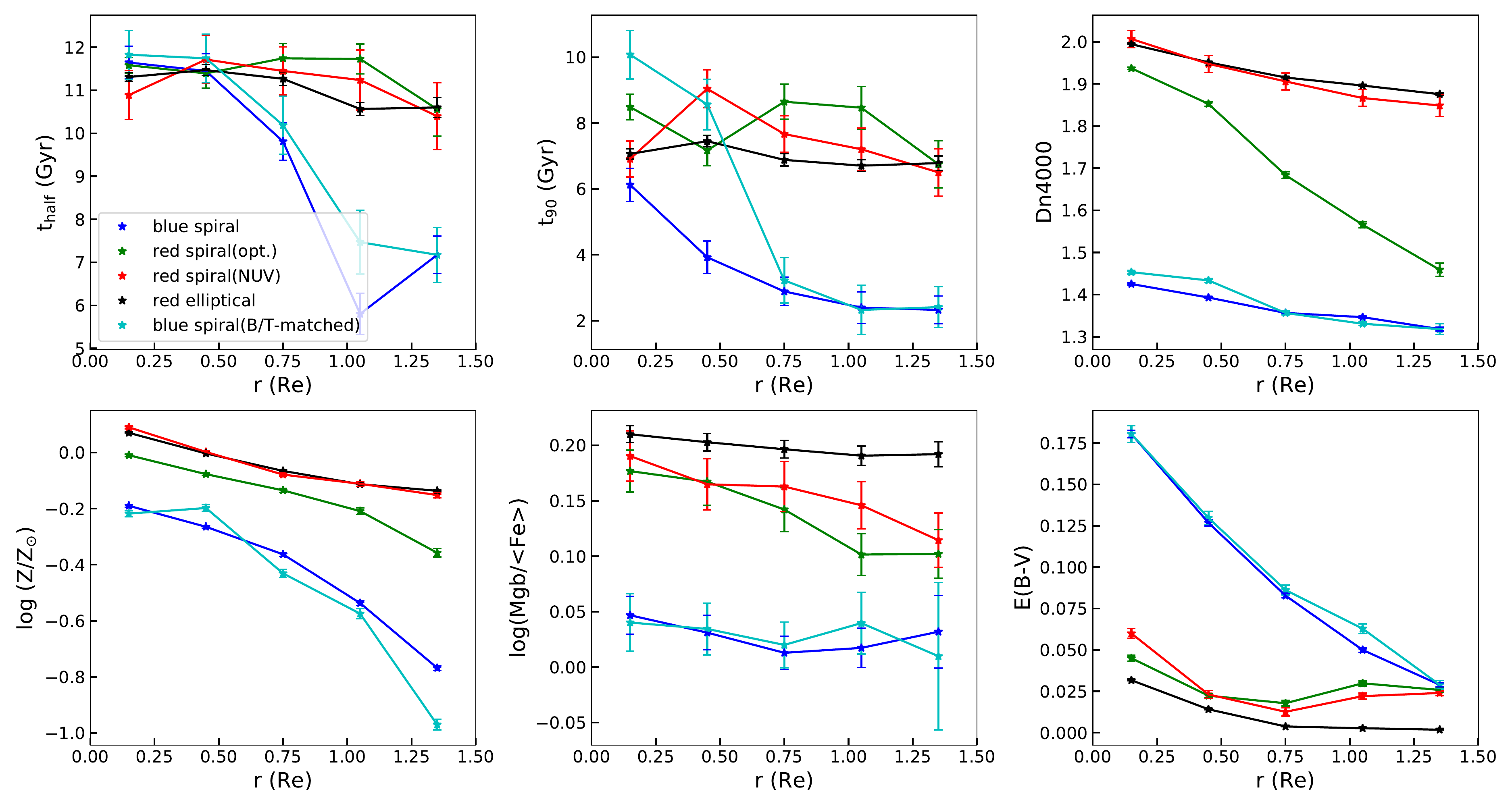}
     \caption{Median radial profiles of half mass formation 
     time $t_{\rm half}$ (top left), 90$\%$ mass formation time $t_{\rm 90}$ (top middle), Dn4000 index (top right),  metallicity (bottom left), \mgbfe (bottom middle), and dust attenuation (bottom right), for red spiral galaxies (green), $NUV$-selected red spiral galaxies (red), blue spiral galaxies (blue), red elliptical galaxies (black) and B/T matched blue spirals(cyan, see \S\ref{ssec:formation}). Results for $t_{\rm half}$, $t_{\rm 90}$ and metallicity are derived from the best-fit model of the stacked spectra in the radial bins, Dn4000 and \mgbfe come from the medians in radial bins of maps from MaNGA DAP, while the dust attenuation are drawn from the map of \cite{Li2020} ( see the example galaxies in \autoref{fig:example}), respectively. In each panel the error bar represents the error of the mean, some error bars are too small to be seen.
     }
    \label{fig:stat}
\end{figure*}

In \autoref{fig:stat} we examine the radial profile of a variety 
of stellar population indicators. These include the half mass formation 
time ($t_{\rm half}$) and 90\% mass formation time ($t_{\rm 90}$), 
estimated from the best-fit stepwise SFH model, as well as 
$Dn4000$ (the spectral break at 4000\AA), stellar metallicity $Z$, 
\mgbfe (the Mgb-to-iron abundance ratio) and $E(B-V)$ (the stellar 
attenuation as quantified by $B-V$ color excess). In each case, we show 
the median profile for each of the four types of galaxies. 
We see that, red spirals (both optical and $NUV$) and red ellipticals
show very similar formation times, in both $t_{\rm half}$ and 
$t_{\rm 90}$ and with fairly flat slopes. These galaxies form 
half of their mass $>10$Gyr ago, and 90\% of their mass $>6$Gyr ago, 
consistent with what we've seen above from their SFHs. Differently
but as expected, blue spirals show strong decrease in the formation 
times as one goes from inner to outer regions. The central region of 
blue spirals shows the same old formation time as the red galaxies, 
with $t_{\rm half}\sim11$Gyr, but slightly smaller $t_{\rm 90}$ 
in the same region which is $\sim6$Gyr compared to $7-8.5$Gyr for 
the other types of galaxies. 

In the top-right panel, as expected, the red ellipticals show 
large $Dn4000$ at all radii, 
with slightly negative gradient, indicative of no recent star 
formation over the whole galaxy area. It is interesting that the 
$Dn4000$ in red spirals selected in $NUV$ closely follows the profile
of red ellipticals, while optically-selected red spirals deviate 
from the $NUV$ counterparts dramatically, showing strong decrease 
in $Dn4000$ with increasing radius. The $Dn4000$ index has been 
commonly used as an indicator of mean stellar age, and in particular 
a value of $Dn4000<1.6$ indicates the existence of young populations 
formed in the past 1-2Gyr \citep[e.g.][]{Kauffmann2003}. 
As can be seen, the $Dn4000$ in optically red spirals drop 
below 1.6 at $\sim R_e$, implying ongoing/recent star formation 
in the outer regions. This is well consistent with the bluer 
images as seen in their $u-r$ color. The use of $Dn4000$ in this 
case provides unique constraints on the recent SFHs of 
our galaxies, thus is able to highlight the difference between 
$NUV$ and optical red spirals which was largely missing in the 
Bayesian inference of the SFHs.

The bottom left panel of \autoref{fig:stat} shows the median 
metallicity profiles of the four types of galaxies. Similar to 
$Dn4000$, the metallicity also shows high similarity 
between $NUV$-selected red spirals and red ellipticals. 
Optically-selected red spirals, however, present $\sim0.1$dex
lower metallicities than ellipticals and $NUV$-selected red spirals
at all radii up to $\sim R_e$, with an even lower metallicity 
at the outermost radius. Blue spirals have much lower metallicities 
at all radii than the other types of galaxies. This result 
again isolates the optically red spirals from the $NUV$-selected 
red spirals and ellipticals. The lower metallicities in the 
optical red spirals but similarly old formation times may 
combine to imply infall of pristine gas into these galaxies
which have provided fuel for recent star formation and lowered 
down the stellar metallicity. 

Moreover, in the bottom middle panel of \autoref{fig:stat} 
we examine the radial profiles of \mgbfe, a proxy 
of the $\alpha$/Fe ratio which is often used to trace timescale 
of the star formation process. The abundance of $\alpha$-elements 
that are produced by core-collapse supernovae can be probed by 
the Mgb index, while Fe5270 and Fe5335 indices trace the abundance 
of Fe generated in type Ia supernova \cite[e.g.][]{Zheng2017}. 
As the explosion of type II supernova happens almost simultaneously
with the star formation events, while low mass stars which are 
progenitors of type Ia supernova need longer time to evolve, 
the relative abundance of $\alpha$-elements and Fe can thus 
characterise the relative importance of violent star burst events 
and low-level continuous star formations. It is seen from the 
figure that red ellipticals have the largest \mgbfe, i.e. being 
$\alpha$-enhanced throughout the galaxy, an effect that has been 
known for decades \citep[e.g.][]{Worthey1992}. This suggests a very 
short timescale for the formation of red ellipticals. The 
$NUV$-selected red spirals and optically-selected red spirals 
have similar \mgbfe, in terms of both amplitudes at given radius 
and the radial slope. At given radius, both types of red spirals 
also show large values of \mgbfe, which are slightly smaller when 
compared to red ellipticals. This indicates similarly short 
(though slightly longer) formation timescales for red spirals, 
compared to red ellipticals. The negative radial gradients in 
red spirals imply a faster formation process in central regions 
than in the outskirt. Blue spirals show lowest \mgbfe\ at 
all radii and little gradient, consistent with the extended 
star formation history as inferred from the spectral fitting with {\tt BIGS}. 

Finally, the bottom right panel of \autoref{fig:stat} shows the 
radial profile of the median dust attenuation parameter $E(B-V)$ 
for the four types of galaxies. As one can see, the two types of 
red spirals and the red ellipticals have very little dust 
attenuation, with $E(B-V)\lesssim 0.05$ at all radii, confirming 
that the red colors of these galaxies are real, but not an 
observational effect of dust attenuation. Blue spirals 
show strongly negative profile in $E(B-V)$, with stronger 
attenuation than the other types of galaxies at all radii 
except the outermost bin. 

In summary, all the radial profiles shown in \autoref{fig:stat} 
point to a simple conclusion that, at least up to 1.5$R_e$, 
both $NUV$ and optical red spirals are similar to red ellipticals 
in the sense that these galaxies form the majority of their 
stars at early times with relatively short formation timescales. 
Once formed, red ellipticals and $NUV$-selected red spirals 
keep quenched, while optically-selected red spirals have 
residual star formation occurring in their outermost regions during 
later times. Blue spirals, in contrast, have an old center 
formed at similarly early times, but have been accumulating 
their disks over longer time-scales, with an extended star 
formation history.

\section{Discussion}
\label{sec:discussion}

\subsection{Difference between optically and $NUV$-selected red spirals}

We find red spirals selected by optical colors and $NUV$-optical colors 
behave differently in terms of present-day stellar populations,
with the former having significantly younger populations in the outskirts
and slightly lower metallicity at fixed radius. This makes the 
optically-selected red spirals fall in the green-valley region or even 
blue-cloud region in some cases in the color-mass diagram (see
\autoref{fig:sample}). The young populations in optically-selected red
spirals imply recent/ongoing star formation, which can be seen from their 
SFHs (the second panel in the top row of \autoref{fig:sfh_nre}). This is 
also consistent with recent works which found a large fraction of atomic 
gas mass in optically-selected passive galaxies \citep[e.g.][]{Zhang2019, 
Guo2020}, as well as earlier works which found optically-selected passive 
spirals are not truly passive, but forming stars at a significant rate 
\citep{Cortese2012}. In contrast, truly passive spirals like the $NUV$-selected
red spirals in our sample are expected to be gas poor, in terms of both 
atomic gas \citep[e.g.][]{Cortese2012,Cortese2020} and molecular gas
\citep[e.g.][]{Luo2020}.
Cross-matching our samples with the ALFALFA-SDSS 
catalogue \citep{Durbala2020}, we found 11 blue spirals, 6 red spirals
(opt.), 2 red spirals (NUV) and 1 red elliptical in our sample 
to have H{\sc i} detections, with a mean H{\sc i}  to stellar mass ratio
of 0.19, 0.15, 0.08 and 0.13 for the four types of galaxies.
Although the sample sizes are rather small, these numbers are in broad
agreement with previous findings that the H{\sc i} detection rate is 
higher in optically selected red spirals compared to truly red 
spirals like the $NUV$-selected sample studied here.

Despite the differences in present-day stellar populations, it is interesting
that the two types of galaxies are very similar in terms of star formation 
history: both formed the majority of stars at early times  (see 
\autoref{fig:sfh_nre}). It is unclear whether the two types of 
galaxies have any evolutionary relationship. $NUV$-selected red spirals 
could be evolutionary remnants of optically red spirals, or optically red 
spirals may be rejuvenated from the $NUV$-selected ones. The latter case
is preferred by our results, considering that the quenching of their star 
formation happened at early times, and that we do not see any residual stellar populations in recent 2 Gyrs in $NUV$-selected red spirals as revealed by the $D4000$ index.

Apparently, the selection scheme has significant impacts on the 
sample properties. Selecting red spiral galaxies with optical colors, 
as done by \cite{Guo2020} and many other studies, may give rise to a sample 
extending to green or even blue colors, thus including star forming 
galaxies with a signficant H{\sc i} gas fraction. In contrast, the 
$NUV$-based selection can reliably select truly red spiral galaxies,
although the optically-red galaxies are interesting by
themselves and deserve more dedicated studies in future.

\subsection{Comparison with previous results}

Due to their red colors but spiral features, the population 
of red spiral galaxies has attracted increasingly more 
attention in the past decade. Most of the previous studies 
have used the single-fiber spectroscopy from SDSS, thus 
limited to the central 1-2kpc of low-z galaxies, although 
the samples were selected by global colors (mostly in the optical). 
For instance, using morphology classifications from the 
Galaxy Zoo project(GZ), \cite{Masters2010} identified 
a sample of red spirals from SDSS according to $g-r$ color, 
and found the red spirals generally have larger $Dn4000$ and 
lower H$\rm \delta_{A}$ compared to blue spirals, 
indicative of reduced recent star formations in red spirals. 
\cite{Tojeiro2013} investigated the star formation histories 
of galaxies of different morphology types, finding recent 
decline of star formation in red spirals. Similar results were 
obtained by \cite{Robaina2012} who found quiescent spirals and 
elliptical galaxies to present similar stellar population 
properties. Our estimates of the SFHs are in good agreement 
with these previous studies, but additionally reveal the 
dominating contribution of the star formation at early times
as well as the radial variation out to large radii. 
At high redshifts, \cite{Bundy2010} studied a sample of 
passive spiral galaxies at $z\sim1-2$ from the COSMOS survey, 
and found the passive spirals to be unlikely the descendants
of star-forming disk galaxies formed at higher redshifts. 
In our work, we also find that the ($NUV$-selected) red spiral 
galaxies resemble red ellipticals in many aspects including the 
SFH, metallicity, $\alpha$-element abundance and stellar 
dust attenuation. On the other hand, blue spirals behave 
quite differently than both red spirals and ellipticals. 
These results strongly imply that massive red spirals 
have followed a distinct evolutionary path when compared to 
their blue counterparts, and that they may share the same 
formation and evolution process(s) with elliptical galaxies. 

Our optically-selected samples are taken from \cite{Guo2020}
who used SDSS data to study the spectroscopic and 
structural parameters in the central region of these galaxies. 
It was found that the massive red spirals have large $Dn4000$
and [Mgb/Fe] in their centers, similarly to massive red ellipticals. 
In addition, the central bulge of massive red spirals followed 
the same $\Sigma_1-M_\ast$ relation as quenched galaxies of similar 
masses, where $\Sigma_1$ is the surface mass density within 
central 1kpc. A large fraction ($\sim$70\%) of the massive 
red spirals had strong bars, ring or shell-like structures, or 
merging features in their image. A similarly high fraction 
of bars has also been found in other studies 
\citep[e.g.][]{Masters2010, Fraser2018}. 
As pointed out in \cite{Guo2020}, these results combined 
suggest that, like massive red ellipticals, the bulge of massive 
red spirals may be formed before $z\sim1-2$ with a short timescale, 
and interactions/mergers might have played an important role. 
In a companion paper, \cite{Hao2019} examined the radial profiles 
of stellar population parameters, including [Mgb/Fe] from MaNGA DAP, 
as well as metallicity and age estimated by \cite{Sanchez2018} 
from the MaNGA spectra by applying the spectral fitting pipeline 
{\tt Pipe3D} \citep{Sanchez2016}. The massive 
red spirals showed profiles in those parameters similar to those 
of massive ellipticals. 

As shown in \autoref{fig:stat}, the main results in \citet{Hao2019} 
are well reproduced using our own spectral fitting pipeline {\tt BIGS}.  
We note that the metallicity gradients in 
\cite{Hao2019} appear shallower than ours for both blue and red spirals. 
Recently, \cite{Lacerna2020} and \cite{Lian2018} also estimated 
metallicity gradients using MaNGA data, respectively for massive 
ellipticals and disk galaxies. Our measurements are consistent with 
theirs. The differences between our measurements and those used in
\cite{Hao2019} may likely be attributed to the different spectral fitting 
pipelines used in deriving the metallicities.
In addition, we have selected a new red spiral sample 
according to $NUV-r$ instead of $u-r$. It is interesting that 
the $NUV$-selected red spirals are almost identical to the red 
ellipticals, not only in radial profiles (also see \autoref{fig:stat})
but also in SFHs (see \autoref{fig:sfh_nre}). This surprisingly 
high degree of similarity reinforces the conclusion that massive 
red spirals, if truly ``red'' (e.g. $NUV-r>5$), indeed share the 
same stellar populations and SFHs as massive red ellipticals. 
Thus, our result implies that the two types of galaxies may share some 
common formation and quenching processes. On the other hand, however, this
implication should not be overemphasized given the limited size of our 
samples.

\subsection{Implications for the formation of massive red spirals}
\label{ssec:formation}
The different SFHs found between massive blue spirals and massive 
red spirals 
in our sample strongly suggest that massive spirals are not formed in 
a simple way. Rather, they are divided into two distinct populations
in terms of formation and evolution paths: massive red spirals that
form at early time with a fast formation process similar to 
the formation process of massive ellipticals, and massive blue spirals 
that form their centers also early but with extended star formation 
occurring in the disk over a long time. One may naively assume 
massive red spirals are evolved remnants of massive blue spirals, 
an idea that 
can be ruled out by the totally different SFHs of the two types of 
galaxies as found in our work. Indeed, the results in \cite{Hao2019} 
and \cite{Guo2020} led to the same conclusion, while the SFHs 
obtained from the Bayesian analysis in the current work make it more 
solid and convincing.

Our analysis has revealed early formation times ($>$10Gyr ago) 
for the stellar populations in both red spirals and ellipticals with $M_\ast>10^{10.5}M_\odot$. Such a high level of similarity strongly
suggests that massive red spirals probably share a common formation 
process with massive ellipticals. In galaxy formation theories, the
structural and kinematic properties of a galaxy is predominantly 
determined by the acquisition and distribution of angular momentum
\citep[e.g.][]{Danovich2015}. Basically, elliptical galaxies form
by losing angular momentum, while disk galaxies form by preserving
and redistributing angular momentum.  In the later case, a disk galaxy
is assumed to form in the host dark matter halo with considerable
fractions of the mass and angular momentum of the halo, and once formed
the disk will grow continuously by converting  gas into stars,
a secular evolution process that extends over a long timescale and
is not associated with mergers \citep[e.g.][]{Fall1980,Mo1998,Dutton2007}.
Massive blue spirals seem to follow this scenario for their extended
SFH in the outer disk. However, the early formation time of their central
region implies that massive blue spirals may also form the majority of 
their stars at the same times as their red counterparts, but continuous 
gas infall and star formation (especially in the disk) at later times 
have made the whole galaxy blue, dusty, and dominated by young and 
metal-poor stars that have relatively small amount of $\alpha$ elements. 
In this case, our results appear to reveal a simple picture for massive 
galaxies as a whole, in which most (if not all) of the massive
galaxies in the local Universe formed the majority of their stars at early
times ($>10$Gyr ago, or $z>2$) with a short formation process, although
the formation process may or may not be the same for different types of 
massive galaxies.

In fact, the formation of disk galaxies at early times has been 
well established with the help of hydro-dynamic simulations, which 
can be mainly divided into two distinct cases. In one case, major mergers
of disk systems may produce both ellipticals or spirals, 
depending on the gas fraction of the progenitors
\citep[e.g.][]{Springel2005,Robertson2006,Hopkins2009,Athanassoula2016,Sparre2017}.
In this case, mergers of disk galaxies with a small gas fraction produce 
elliptical galaxies. In case of gas-rich mergers, a rotating 
bulge is formed in the first place by rapid gas infall and star formation
as driven by tidal torques, followed by the formation of a star-forming
disk through cooling and settling of the remaining gas, which
typically takes a timescale of $\sim1$Gyr \citep[e.g.][]{Springel2005}.
The slightly negative [Mgb/Fe] gradient as seen in the massive spirals
in our sample supports a later formation time of the disk compared to
the central region (see the bottom-middle panel in \autoref{fig:stat}).
Gas-rich mergers are expected to happen frequently only at high redshifts
when cold gas is substantially locked in/around galaxies. The formation
redshift of $z=2$ or higher as suggested by our analysis is known to be
the cosmic epoch with highest densities of star formation rate and
black hole accretion  rate, which are largely triggered by galaxy-galaxy mergers.

On the other hand, however, simulations have also suggested a non-merger
origin of elliptical and red spiral galaxies at high redshift
\citep[e.g.][]{Dekel2014,Zolotov2015}. In this case, a gas-rich, star-forming but 
highly perturbed disc is formed in the first place, and the dissipative,
quick compaction of the gas disc leads to the formation of a compact, 
blue ``nugget'', which is quickly converted into a compact, red ``nugget''
due to fast quenching of star formation. Later on, the galaxy may 
gradually grow and extend by dry mergers, possibly developing a new disc 
or a surrounding ring-like structure. The galaxy formed this way should 
have an old center and a late-type morphology, with either red or blue colors
depending on how {\em dry} the gradual growth process is. This formation 
process may be applicable to the massive red spirals in our sample, as 
long as the outer disk is quenched also quickly so as to have similar SFHs 
across the whole galaxy. In \citet{Zolotov2015}, however, although high-mass 
galaxies tend to compactify and quench rather efficiently, their overall 
SFR quenching rate within 10kpc is slower than the inner quenching 
during the post-compaction phase. This appears to imply that the picture of 
compaction formation is unlikely the best model for massive red spirals. 
On the other hand, however, as one can see from \autoref{fig:sfh_nre}, 
the SFH indeed vary from radius to radius, and the outer regions get fully 
quenched later than the central region by a few Gyr. Therefore, there is 
still room for the compaction model and measurements of SFH with 
better time resolution would be needed if one were to discriminate the 
model. 

\citet{Guo2020} suggested that massive ellipticals were formed through  
compaction of gas-rich disks, while massive red spirals were formed by 
gas-rich major mergers. The progenitors in both processes are gas-rich disk 
galaxies which dominate the galaxy population at high redshifts. 
Therefore, the scenario suggested by \citet{Guo2020} provides a 
plausible picture, considering our result that both types of galaxies
were formed at early times. As discussed above, however, 
ellipticals could also be formed by relatively gas-poor major mergers, while 
red spirals could also be formed by compaction of gas-rich disks followed 
by growth of a surrounding disk during the post-compaction phase. These cases
may not play dominating roles, but none of them can be simply ruled out 
by current data. It is thus likely that, the different formation mechanisms 
work together, but contributing to varying degrees depending on the 
detailed properties of the progenitor galaxies (e.g. gas content, angular 
momentum, merger rate, etc.). More works are needed in future to better 
determine the relative roles of different mechanisms for different types 
of galaxies.

Our SFH measurements have revealed little evolution in both ellipticals
and red spirals in the past $\sim6$Gyr, and relatively weak evolution
at intermediate times ($6-10$Gyr ago, see \autoref{fig:sfh_nre}).
It is natural to ask how those red spirals have been able to keep
themselves quenched ever since formation. In the literature a 
number of quenching mechanisms have been suggested, such as morphology 
quenching as originally proposed by \citet{Martig2009}, bar quenching 
as recently studied in detail with numerical simulations 
\citep[e.g.][]{Spinoso2017,Khoperskov2018}, quenching due to high angular
momentum of infalling gas \citep[e.g.][]{Peng2020}, and AGN feedback 
\citep[e.g.][]{Luo2020}. It is beyond the scope and capability of our work 
to discriminate these mechanisms, but we have done a simple test on 
the mechanism of morphology quenching. We select a subset of the 
blue spirals in our sample by requiring the distribution of bulge-to-total 
stellar mass ratio ($B/T$) to be closely matched with that of the optically
selected red spirals, and we show the radial profiles of stellar population
properties of this subsample as cyan lines/symbols in \autoref{fig:stat}.
As one can see, when matched in $B/T$ with the red spirals, the blue spirals
remain unchanged in all the profiles except the profile of $t_{90}$
(the 90\%-mass formation time), which becomes significantly steeper within
$\sim0.5$Re, with early formation times in the galactic center that are
comparable to the red galaxies. This indicates that the early formation
times found in the center of blue spirals as a whole are largely contributed
by the subset of galaxies with early-formed bulges. The different
formation times and populations in the outer regions between blue and
red spirals, which are held even when $B/T$ is closely matched,
imply that the central bulge is unlikely to take effect in all systems.
On the other hand, however, the median result presented here can not 
rule out the possibility that some of those systems have been affected by 
their central bulges. Obviously, more investigations are needed, both 
observationally and theoretically, to fully understand the quenching 
mechanism for massive red spiral galaxies.

Finally, environment effects cannot be ignored when studying
galaxy evolution. For red spirals, \cite{Masters2010} found that the 
fraction of red spirals peaks in intermediate density regimes, but 
environment alone is not enough to make the galaxies red. Here, we 
have made use of the Galaxy Environment for MaNGA Value Added 
Catalog (GEMA-VAC,\citealt{Argudo2015}) to examine the environment of 
our sample galaxies. For each galaxy in the MaNGA, the GEMA-VAC adopts 
the SDSS galaxy group catalogues from \cite{yang2007} to classify 
the galaxy as either the central and a satellite galaxy in its host 
group, and also provides the average local density within 1Mpc 
of the galaxy. We find the different types of galaxies in our sample
to show similar central/satellite fractions, as well as similar 
dentities of the local environment. This implies that environment 
is unlikely to play important roles in regulating the SFHs of our 
galaxies. However, again, this result should not be overemphasized
due to the relatively small size of our samples. More works with 
larger samples are needed in future.

\section{Summary}
\label{sec:summary}

In this paper we investigate the SFH of massive red spiral galaxies with 
stellar mass $M_\ast>10^{10.5}M_{\odot}$ by analyzing the MaNGA spectra 
with our Bayesian inference code {\tt BIGS}. We consider two sample of red spirals,
selected by color index of $u-r$ or $NUV-r$, as well as comparison samples of 
blue spirals and red ellipticals that have similar masses to the red spiral
galaxies. The spatially resolved spectra from MaNGA of each 
galaxy are divided into five radial bins, and those in each radius are stacked 
to achieve high enough SNR. We apply {\tt BIGS} to fit the staked spectra,
using three different functional forms to model the star formation history, 
as well as constraining their present-day stellar population properties.

Our main results can be summarized as follows:
\begin{itemize}

\item
Our best-fit SFHs reveal that massive red spirals and red ellipticals have 
very similar star formation histories : they formed more than half of their 
stellar mass at least 10 Gyr ago and more than 90\% of their stellar masses 
at least 6 Gyr ago. $NUV$-selected red spirals and red ellipticals quenched 
almost entirely since then, but residual star formations are seen in the 
outer regions of optically selected red spirals. The centres of blue spirals 
also already formed before 10 Gyr ago, but the outer disk formed significantly 
latter through extended star formation over a long timescale.  

\item
The derived SFHs are robust against the variation of the assumed SFH models.
Using Bayesian model selection approaches, we confirm that the star 
formation history of massive red spirals and massive red ellipticals can be 
characterised by a simple $\Gamma$ model, which indicates that they have 
experienced only one major star formation episode. For massive blue spirals,
in contrast, the $\Gamma$ model behaves significantly worse than the $\Gamma$+B 
and stepwise models and give a very biased result, indicating more complex SFHs
in those galaxies.

\item
SFHs constrained from spectral fitting are in good agreement with those 
indicated by spectra features. High $D4000$ values and weak/no H$\alpha$ 
emission are found in both $NUV$-selected red spirals and red ellipticals,
indicating that both types of galaxies have quenched their star formations 
throughout the galaxy. Optically selected red spirals have similarly high 
$D4000$ at galactic centres, but relatively low $D4000$ associated with detectable 
H$\alpha$ emissions in the outer regions, suggesting a resident level of 
star formations. Massive blue spirals present low $D4000$ and strong H$\alpha$ 
emission at all radii, indicating ongoing/recent star formation. High 
\mgbfe ~values are found in both red ellipticals and red spirals, consistent
with their early formation and fast quenching, while the opposites are found 
for blue spirals.

\item
Our results clearly show that the majority of massive red 
spirals, especially those selected by their $NUV-r$ color, are  
unlikely evolutionary remnants of massive blue spirals due to the 
completely different formation times and SFHs. The similar SFHs and 
stellar populations in massive red spirals
and ellipticals imply that the stellar contents in those systems 
could have formed through similar processes, but more investigations 
with larger samples are needed in order to pin down the exact 
formation and evolution paths.

\end{itemize}

The success of our analysis relies on both the Bayesian inference
of full spectral fitting and the integral field spectroscopy from MaNGA. 
However, as pointed out several times, the relatively small samples 
should be kept in mind when interpretating our results. Due to the small
sample size, some rare but important populations may be under-represented
in our samples, and environment- and gas-related quenching processes could not
be examined in depth. We expect to significantly enlarge our samples by 
using the full MaNGA sample which will include 10,000 galaxies, thus more 
than two times larger than the DR15 samples used in the current work.
In addition, single-fiber spectroscopy from even larger surveys such as
SDSS may also help to validate our conclusions, at least on the SFH of 
the central region of galaxies. With much larger samples, one will be able 
to additionally take into account the environment and H{\sc I} gas content, 
thus disentangle the relative roles of all potential effects. 
Moreover, next-generate surveys will provide large 
samples of spectra for studies of SFH of galaxies at $z>1$, which will 
enble the early star formation and quenching phases of the massive red 
spirals to be better diagnosed. In fact, \cite{Carnall2019} have made a 
step forward along this line by studying a sample of massive quiescent 
galaxies at $1.0 < z < 1.3$ observed by the VANDELS 
survey \citep{ McLure2018}. 
Some of those early quiescent galaxies are found to form in extreme 
starbursts and quenched at $z<2$, consistent with the merger origin 
as discussed above.

\acknowledgments
This work is supported by the National Key R\&D Program of China
(grant No. 2018YFA0404502), and the National Science 
Foundation of China (grant Nos. 11821303, 11733002, 11973030, 
11673015, 11733004, 11761131004, 11761141012).

Funding for SDSS-IV has been provided by the Alfred P. Sloan Foundation and 
Participating Institutions. Additional funding towards SDSS-IV has been provided 
by the US Department of Energy Office of Science. SDSS-IV acknowledges support 
and resources from the Centre for High-Performance Computing at the University 
of Utah. The SDSS web site is www.sdss.org.

SDSS-IV is managed by the Astrophysical Research Consortium for the Participating 
Institutions of the SDSS Collaboration including the Brazilian Participation Group, 
the Carnegie Institution for Science, Carnegie Mellon University, the Chilean 
Participation Group, the French Participation Group, Harvard–Smithsonian Center 
for Astrophysics, Instituto de Astrofsica de Canarias, The Johns Hopkins University, 
Kavli Institute for the Physics and Mathematics of the Universe (IPMU)/University 
of Tokyo, Lawrence Berkeley National Laboratory, Leibniz Institut fur Astrophysik 
Potsdam (AIP), Max-Planck-Institut fur Astronomie (MPIA Hei- delberg), 
Max-Planck-Institut fur Astrophysik (MPA Garching), Max-Planck-Institut fur 
Extraterrestrische Physik (MPE), National Astronomical Observatory of China, 
New Mexico State University, New York University, University of Notre Dame, 
Observatario Nacional/MCTI, The Ohio State University, Pennsylvania State University, 
Shanghai Astronomical Observatory, United Kingdom Participation Group, Universidad 
Nacional Autonoma de Mexico, University of Arizona, University of Colorado Boulder, 
University of Oxford, University of Portsmouth, University of Utah, University of 
Virginia, University of Washington, University of Wisconsin, Vanderbilt University 
and Yale University.

\bibliographystyle{aasjournal}

\bibliography{szhou}

\end{document}